\documentclass[useAMS,usenatbib]{mn2e}
\usepackage{graphicx}

\def\mjup{{\rm M}_{2\!\!{\!\hspace{0.01in}\hbox{\rm \tiny \bf +}}}}
\def\rmscr#1{{\hbox{\rm \scriptsize #1}}}

\def\mjup{{\rm M}_{\rm J}}
\def\msun{{\rm M}_\odot}

\begin{document}
\title[Orbital Evolution]{Orbital evolution with white-dwarf kicks}

\author[J. Heyl]{Jeremy Heyl$^{1}$\\
$^{1}$Department of Physics and Astronomy, University of British Columbia, Vancouver, British Columbia, Canada, V6T 1Z1 \\
Email: heyl@phas.ubc.ca; Canada Research Chair}

\date{\today}

\pagerange{\pageref{firstpage}--\pageref{lastpage}} \pubyear{2007}

\maketitle

\label{firstpage}

\begin{abstract}
  Recent observations of white dwarfs in globular clusters indicate
  that these stars may get a velocity kick during their time as
  giants.  This velocity kick could originate naturally if the mass
  loss while on the asymptotic giant branch is slightly asymmetric.
  If white dwarfs get a kick comparable to the orbital velocity of the
  binary, the initial Runge-Lenz vector (eccentricity vector) of the
  orbit is damped to be replaced by a component pointing toward the
  cross product of the initial angular momentum and the force.  The
  final eccentricity may be of order unity and if the kick is
  sufficiently large, the system may be disrupted.  These results may
  have important ramifications for the evolution of binary stars and
  planetary systems.
\end{abstract}
\begin{keywords}
white dwarfs --- stars : AGB and post-AGB --- binaries -- stars: mass loss --- stars: winds, outflows 
\end{keywords}

\section{Introduction}
\label{sec:introduction}

\citet{1998A&A...333..603S} proposed that white dwarfs can acquire
their observed rotation rates from mild kicks generated by asymmetric
winds toward the end of their time on the asymptotic giant branch
(AGB) \citep{1993ApJ...413..641V}.  \citet{2003ApJ...595L..53F}
invoked these mild kicks to explain a putative dearth of white dwarfs
in open clusters
\citep[e.g.][]{1977A&A....59..411W,2001AJ....122.3239K}.  The expected
signature of white dwarf kicks has been observed in M4 and NGC~6397
\citep{2007Davis}.  They found that young white dwarfs are less
centrally concentrated than either their progenitors near the top of
the main sequence or older white dwarfs whose velocity distribution
has had a chance to relax.

If white-dwarf kicks can have dramatic effects on the distribution of
stars in a globular cluster, perhaps they could also affect other
bound systems such as binary stars.   The case of binary stars is
interesting for a second reason as they represent a different physical
regime than either white-dwarf kicks in clusters or neutron-star kicks
in binaries.  Whereas the latter two effects are impulsive, the white
dwarf kick accumulates over many orbital periods, so its secular
effect is more subtle to calculate.

\section{Calculations}
\label{sec:calculations}

The fate and evolution of binaries with asymmetric mass loss is
complicated.  Specifically the mass loss turns on and off over many
binary orbital periods so its influence on the binary orbit may be
adiabatic (in contrast with mass loss during the formation of a
neutron star).  If the mass loss is symmetric the orbit increases in
size in inverse proportion to the decreasing mass of the system.
Specifically, the angular momentum of the orbit, the Runge-Lenz
vector, and the product of the binding energy and the period of the
orbit are adiabatic invariants.

Including the effects of an asymmetric wind requires further analysis.
The total kick imparted by the wind is on the order of a few to ten
kilometers per second \citep{2007Davis} over many tens of thousands of
years, so the force exerted by the kick is typically much smaller than
the gravitational forces in the binary that result in accelerations
of tens of kilometers per second per year --- the kick is a
perturbation.  Even in the limit that the kick can not be treated
perturbatively but it is still adiabatic, one can use the fact that
the Kepler problem with a constant external force is tractable in
parabolic coordinates to find that the angular momentum and the
Runge-Lenz vector along the direction of the kick and the product of
the total mass of the system and the semimajor axis are adiabatic
invariants.  However, these three quantities are insufficient to
determine the final state of the binary, most importantly, its
eccentricity.

\subsection{Centre of mass motion}
It is useful to separate the force that imparts the kick into a force
on the centre of mass and a torque about the centre of mass.  Using
centre of mass coordinates, the Lagrangian of the binary including the
kick on the giant star is
\begin{equation}
L = \frac{1}{2} M \dot R^2 + \frac{1}{2} \mu \dot r^2 +
\frac{GM\mu}{r} + {\bf F}_1 \cdot \left ( {\bf R} + \frac{M_2}{M} {\bf
    r} \right )
\label{eq:1}
\end{equation}
where $M$ is the total mass, $\mu$ is the reduced mass, ${\bf R}$ is
the position of the centre of mass and ${\bf r}={\bf r}_1-{\bf r}_2$.
The force exerted on a single giant star by its asymmetric wind is
\begin{equation}
{\bf F}_1 = {\bf b} v_\rmscr{wind} \left | \dot M \right |
\label{eq:2}
\end{equation}
where ${\bf b}$ is a dimensionless vector characterising the
asymmetry; the magnitude of ${\bf b}$ can range from zero for a
symmetric wind to unity for a wind that blows only along a single
direction lik a rocket.  Therefore, the acceleration of the star is given by
\begin{equation}
\frac{d {\bf v}_1}{dt} = \frac{1}{M_1} {\bf b} v_\rmscr{wind} \left | \dot M \right |
\label{eq:3}
\end{equation}
and the total velocity kick for the single star is
\begin{equation}
\Delta {\bf v}_\rmscr{single} = 
 {\bf b} v_\rmscr{wind} \ln \frac {M_{WD,1}}{M_1}
\label{eq:4}
\end{equation}
so the force on the single star can be written in terms of the
resulting kick
\begin{equation}
{\bf F}_1 = \frac{\Delta {\bf v}_\rmscr{single}}{\ln \left (M_{WD,1}/M_1\right )}
\left | \dot M \right |.
\label{eq:5}
\end{equation}
From Eq.~(\ref{eq:1}), the equation of motion for ${\bf R}$ is
\begin{equation}
{\bf F}_1 = \left ( M_1 + M_2 \right ) \ddot {\bf R} 
\label{eq:6}
\end{equation}
and
\begin{equation}
\frac{d {\bf v}_\rmscr{binary}}{dt} = 
\frac{1}{M_1 + M_2} 
\frac{\Delta {\bf v}_\rmscr{single}}{\ln \left (M_{WD,1}/M_1\right )}
\left | \dot M \right |
\label{eq:7}
\end{equation}
so the 
total kick to the centre of mass is 
\begin{equation}
\Delta v^{(1)}_\rmscr{binary} = 
\frac{\Delta v_\rmscr{single}(M_1) }{\ln \left (M_{WD,1}/M_1\right )}
\ln \frac{M_{WD,1}+M_2}{M_1+M_2} 
\label{eq:8}
\end{equation}
where the product of the velocity and asymmetry of the wind is assumed
to be constant (the mass-loss rate may vary) as the mass of the giant
star decreases and $\Delta v_\rmscr{single}(M_1)$ is the typical
velocity kick imparted to a main-sequence star of mass $M_1$.  As the
secondary becomes a white dwarf, the binary receives a second kick
\begin{equation}
  \Delta v^{(2)}_\rmscr{binary} = \frac{\Delta v_\rmscr{single}(M_2) }{\ln \left (M_{WD,2}/M_2\right)}
  \ln \frac{M_{WD,1}+M_{WD,2}}{M_{WD,1}+M_2}.
\label{eq:9}
\end{equation}
If the mass of the primary and that of the secondary are about equal
\citep[due to dynamical biasing, for example,][]{1993MNRAS.262..800M},
the direct sum of the two kicks would equal the kick received by a
single star.  However, the two kicks will generally not be aligned
with each other, so they must be added in quadrature yielding a
smaller combined kick for the binary.  The total change of the
velocity of the centre of mass of the binary is typically about
70\%-80\% of kick imparted to an individual star \citep[using the
initial-final mass relation of][]{1983ARA&A..21..271I}.

\subsection{Constant Masses}

To treat the evolution of the orbit, the influence of the wind on the
orbital dynamics must be determined, especially whether or not it is
adiabatic.  Fortunately, for the reasons mentioned earlier, the
dynamics of the orbit with the asymmetric wind are remarkably similar
to the Kepler problem. \citet{1964PhRv..133.1352R} found a
generalisation of the Runge-Lenz vector in the presence of an external
force (${\bf F}={\bf F}_1$).  Specifically, the generalised Runge-Lenz vector is
given by
\begin{equation}
{\bf C} = {\hat {\bf r}} + \frac{{\bf L} \!\times\! {\bf p}}{GM\mu^2} - \frac{\left({\bf r}\!\times\! {\bf F}\right)\!\times\! {\bf r}}{2GM\mu}
\label{eq:10}
\end{equation}
where $M$ is the total mass of the binary, $\mu$ is the reduced mass
of the binary, ${\bf L}$ is the orbital angular momentum, ${\bf r}$ is
the relative position of the two stars and ${\bf p}$ is their relative
momentum.  The first two terms give the Keplerian Runge-Lenz vector.
In this system only the component of ${\bf C}$ parallel to the applied
force is conserved.  The vector evolves according to
\begin{equation}
\dot {\bf C} = \frac{3}{2GM\mu^2} {\bf L}\!\times\! {\bf F}.
\label{eq:11}
\end{equation}
The generalised Runge-Lenz vector is no longer perpendicular to the
angular momentum, 
\begin{equation}
{\bf L}\cdot {\bf C} = \frac{r^2 {\bf F}\cdot{\bf L}}{2GM\mu}
\label{eq:12}
\end{equation}
and the relative position of the two bodies is constrained by
\begin{equation}
{\bf C}\cdot {\bf r} = r - \frac{\bf L^2}{GM\mu^2} = r - l
\label{eq:13}
\end{equation}
where $l$ is the semilatus rectum.  With some rearrangement this
yields the equation for an ellipse with a focus at the origin in polar
coordinates
\begin{equation}
r = \frac{l}{1 - \left | {\bf C} \right | \cos \theta } = \frac{l}{1+e\cos\theta}
\label{eq:14}
\end{equation}
where $\theta$ is the angle between ${\bf r}$ and ${\bf C}$.  This
equation shows that the vector ${\bf C}$ has a length equal to the
eccentricity of the orbit and points from the pericenter toward the
apocentre of the orbit.  The external force naturally exerts a torque
on the orbit, so the angular momentum is not conserved,
\begin{equation}
\dot {\bf L} = {\bf r} \times {\bf F}.
\label{eq:15}
\end{equation}
However, because the external force is taken to be a perturbation,
both ${\bf C}$ and ${\bf L}$ can be taken to be constant over an
orbital period ($P$).  When averaged over an orbit the only component
of $\dot {\bf L}$ that remains must be perpendicular to both ${\bf F}$
and ${\bf C}$ because the secondary spends the same amount of time
between pericentre and apocentre as between apocentre and pericentre
and consequently the component of ${\bf r}$ perpendicular to ${\bf C}$
vanishes.  The remaining component of the torque is
\begin{equation}
\dot {\bf L}_{{\bf C}\times {\bf F}} = \frac{{\bf C}\cdot  {\bf r}}{{\bf C} \cdot {\bf C}}
{\bf C} \times {\bf F} = \frac{l}{e^2} \left ( \frac{1}{1+e\cos\theta}
  - 1 \right ) {\bf C} \times {\bf F}.
\label{eq:16}
\end{equation}
yielding the torque averaged over an orbit
\begin{eqnarray}
  \left \langle \dot {\bf L} \right \rangle &=&  {\bf C} \times {\bf
    F}  \left [
\frac{l}{Pe^2} \int_0^P \left ( \frac{1}{1+e\cos\theta} 
  - 1 \right ) dt\right ]
\label{eq:17} \\
&=& {\bf C} \!\times\! {\bf F} \left [ 
\frac{l}{Pe^2} \int_0^{2\pi} \frac{1}{1+e\cos\theta} 
\frac{d\theta}{\dot \theta} - \frac{l}{e^2} \right ] 
\label{eq:18}
\end{eqnarray}
Using the definition of the angular momentum, $L=\mu r^2 \dot \theta$ gives
\begin{eqnarray}
  \left \langle \dot {\bf L} \right \rangle
&=& {\bf C} \!\times\! {\bf F} \left [ 
\frac{l^3}{Pe^2} \frac{\mu}{L} 
\int_0^{2\pi} \frac{1}{\left ( 1+e\cos\theta \right )^3} 
d\theta - \frac{l}{e^2}
 \right ].
\label{eq:19}
\end{eqnarray}
To calculate this integral, the substitution $z=e^{i\theta}$ converts
it to a contour integral, so
\begin{equation}
\cos\theta=\frac{1}{2} \left (z + \frac{1}{z}\right ), d \theta
=\frac{dz}{iz}
\label{eq:20}
\end{equation}
which yields
\begin{eqnarray}
  \left \langle \dot {\bf L} \right \rangle
&=& {\bf C} \!\times\! {\bf F} \left [ 
\frac{1}{e^2} \frac{\mu l^3}{LP} 
\oint \frac{1}{\left ( 1+\frac{ez}{2} + \frac{e}{2z} \right )^3} 
\frac{dz}{iz} - \frac{l}{e^2}
 \right ]
\label{eq:21} \\
&=&
 {\bf C} \!\times\! {\bf F} \left [ 
-\frac{8i}{e^5}  \frac{\mu l^3}{LP} 
\oint \frac{z^2}{\left ( z^2 + \frac{2z}{e} + 1 \right )^3} dz - \frac{l}{e^2}
 \right ] 
\label{eq:22} \\
&=&
 {\bf C} \!\times\! {\bf F} \left [ 
-\frac{8i}{e^5}  \frac{\mu l^3}{LP} 
\oint \frac{z^2 dz}{\left ( z - z_+ \right )^3 \left (z-z_-\right)^3} - \frac{l}{e^2}
 \right ] 
\label{eq:23}
\end{eqnarray}
where the contour is the unit circle and 
\begin{equation}
z_\pm = \frac{1}{e} \left ( \pm\sqrt{1-e^2} - 1 \right ),
\label{eq:24}
\end{equation}
so $z=z_+$ is a pole within the contour.  Using the Cauchy integral
formula yields after some algebraic simplifications
\begin{equation}
  \left \langle \dot {\bf L} \right \rangle =
 {\bf C} \!\times\! {\bf F} \left [ 
 \frac{\mu l^3}{LP} \frac{\pi \left ( 2+e^2 \right )}{e^2\left ( 1 - e^2\right )^{5/2}}
 - \frac{l}{e^2}
 \right ] .
\label{eq:25}
\end{equation}
Using the value of the semilatus rectum from Eq.~(\ref{eq:13}) and
Kepler's third law, $P=\sqrt{GM/a^3}/(2\pi)$ where $a=l/(1-e^2)$ is the semimajor
axis, yields
\begin{equation}
  \left \langle \dot {\bf L} \right \rangle =
 {\bf C} \!\times\! {\bf F} \left [ 
 \frac{l^{5/2}}{a^{3/2}} \frac{ 2+e^2 }{2e^2\left ( 1 - e^2\right )^{5/2}}
 - \frac{l}{e^2} 
 \right ] =
\frac{3a}{2} {\bf C} \!\times\! {\bf F}.
\label{eq:26}
\end{equation}
To estimate
the timescale of the evolution of the orbital parameters, Eqs.~(\ref{eq:11})
and~(\ref{eq:26}) can be combined to yield
\begin{equation}
\ddot {\bf C} = \frac{9}{4} \frac{a}{GM\mu^2} \left ( {\bf C} \!\times\! {\bf F}\right ) \!\times\! {\bf F} 
,~~\ddot {\bf L} = \frac{9}{4} \frac{a}{GM\mu^2} \left ( {\bf L} \!\times\! {\bf F}\right ) \!\times\! {\bf F} 
\label{eq:27}
\end{equation} 
if the masses of the stars and the force are taken to be constant.  If
the prefactor is constant, these are the equations for a harmonic
oscillator with a frequency of zero along the direction of ${\bf F}$
and a frequency of 
\begin{equation}
\omega=\frac{3}{2}\frac{F}{\mu} \sqrt{\frac{a}{GM}} 
\label{eq:51}
\end{equation}
perpendicular to
${\bf F}$.  The evolutionary timescale is
\begin{equation}
\tau  = \frac{1}{\omega} =  \frac{2}{3}\frac{\mu}{F} \sqrt{\frac{GM}{a}} 
\approx \Delta t_{\rm wind} \frac{v_{\rm orbital}}{\Delta v_{\rm single}}
\label{eq:28}
\end{equation}
When the prefactor can be taken to be constant,
Eq.~(\ref{eq:27}) can be solved by splitting the Runge-Lenz vector into
a component parallel and perpendicular to the applied force, yielding
\begin{eqnarray}
C_{\|} &=& {\rm Constant}, \label{eq:29}\\
{\bf C}_\perp &=& {\bf C}_{\perp,0} \cos \omega t + \sqrt{1-e_0^2} \frac{{\bf L_0}\!\times\!{\bf F}}{L_0F} \sin \omega t .
\label{eq:30}
\end{eqnarray}
The angular momentum has a similar solution
\begin{eqnarray}
L_{\|} &=& {\rm Constant}, 
\label{eq:31}\\
{\bf L}_\perp &=& {\bf L}_{\perp,0} \cos \omega t + \frac{L_0}{\sqrt{1-e_0^2}} \frac{{\bf C_0}\!\times\!{\bf F}}{F} \sin \omega t .
\label{eq:32}
\end{eqnarray}
In both cases the component of the vector along the direction of the
force is conserved, and the perpendicular components vary harmonically.

\subsection{Variable Masses}

In general the masses in the binary and the force are not constant, so
Eq.~(\ref{eq:11}) must be modified to yield
\begin{eqnarray}
\dot {\bf C} &=& \frac{3}{2GM\mu^2} {\bf L} \!\times\! {\bf F}  -\frac{\left ({\bf r} \!\times\! \dot {\bf F}\right) \!\times\! \bf r}{2GM\mu}
\nonumber \\*
& & ~~~
+ \frac{d}{dt} \left ( \frac{1}{GM\mu^2} \right ) \left [ {\bf L}\!\times\! {\bf p} - \frac{\mu}{2} \left ( {\bf r}\!\times\! {\bf F} \right ) \!\times\! {\bf r} \right ] 
\label{eq:33}
\end{eqnarray}
If the applied force is written as ${\bf F}=(M_2/M) {\bf b} v_{\rm
  wind}|\dot M|$ where ${\bf b}$ is a dimensionless vector
characterising the asymmetry in the wind and the mass ratio gives the
force in the centre of mass frame from Eq.~\ref{eq:1},
Eq.~(\ref{eq:33}) can be expanded in powers of $\dot M$.  If $v_{\rm
  wind}$ and ${\bf b}$ are taken to be functions of the mass of the
star, the first and third terms are first order in $\dot M$, and the
second and fourth terms are second order in $\dot M$.  The orbit
average of the third term vanishes because the orbit is closed to
lowest order in $\dot M$.  This leaves only the first term or
Eq.~(\ref{eq:11}) at lowest order in $\dot M$.  In taking the orbit
average to obtain Eq.~(\ref{eq:26}), it was assumed that ${\bf L}$ and
${\bf C}$ were constant in time; therefore Eq.~(\ref{eq:26}) is only
accurate to first order in $\dot M$, so it is consistent at lowest
order to use Eq.~(\ref{eq:11}) and~(\ref{eq:26}).

In general the prefactor is not constant; however the combination $Ma$
is an adiabatic invariant because the equations of motion of the
binary are separable in cylindrical coordinates as discussed in
\S\ref{sec:introduction}.  Furthermore, the applied force is
proportional to $\dot M$, so Eq.~(\ref{eq:27}) can be written with the
total mass of the system, $M$, as the dependent variable.  This yields
\begin{equation}
\frac{d^2}{dM^2} {\bf C} = \frac{9}{4}  \frac{1}{M^2 M_1^2}
\frac{Ma}{G} v^2_{\rm wind} \left ( {\bf C} \!\times\! {\bf b}\right ) \!\times\! {\bf b}
\label{eq:34}
\end{equation} 
and similarly for the angular momentum.  This assumes that the value
of $\dot M$ is constant in time.  One can argue that the mass-loss
rate depends on the mass of the star, so changes in the mass-loss rate
would only appear at higher order.

This equation may be rewritten by changing variables to
\begin{equation}
x=\ln \left ( 1 + \frac{M_2}{M_1} \right )
\label{eq:35}\end{equation}
where star 1 is the star that loses mass and the variable $x$
increases in time.  This yields, 
\begin{equation}
\frac{d^2}{dx^2} {\bf C} + \left(1+2\frac{M_1}{M_2}\right) \frac{d}{dx} {\bf C} = \frac{9}{4}  \frac{1}{M_2^2} \frac{Ma}{G} v^2_{\rm wind} \left ( {\bf C}\!\times\!{\bf b}\right )\!\times\!{\bf b},
\label{eq:36}
\end{equation} 
the equation for a damped harmonic oscillator with a damping parameter
that decreases in time. The original Eq.~\ref{eq:34} yields the solution
\begin{eqnarray}
{\bf C}_\perp(M_1) &=& {\bf C}_1 \left(\frac{M}{M_0}\right)^{\frac{1}{2}-\alpha}
\left(\frac{M_1}{M_{1,0}} \right)^{\frac{1}{2}+\alpha} \nonumber \\*
& & ~~~
+ {\bf C}_2 \left(\frac{M}{M_0}\right)^{\frac{1}{2}+\alpha}
\left(\frac{M_1}{M_{1,0}} \right)^{\frac{1}{2}-\alpha}
\label{eq:37}
\end{eqnarray}
where
\begin{equation}
\alpha = \frac{1}{2} \sqrt { 1 - 4\beta_0^2}
\label{eq:38}
\end{equation}
and the undamped frequency of the oscillator is 
\begin{equation}
\beta_0 = \frac{3}{2} b v_{\rm wind} \sqrt{\frac{Ma}{GM_2^2}}
 \approx 3 \frac{\Delta v_{\rm
    single}}{v_{\rm orbital}}
  \label{eq:39}
\end{equation}
and the parallel component is constant as before.  This frequency is
dimensionless because the eccentricity is now written as a
dimensionless function of
the mass ratios rather than time.

The initial conditions give
\begin{eqnarray}
{\bf C}_{1,2} &=& 
\frac{1}{2} \left ( 1 \mp \frac{M_0 + M_{1,0}}{\sqrt{1-4\beta_0^2} M_2} \right )
{\bf C}_{\perp,0} \nonumber \\*
& & ~~~ 
 \pm  \beta_0 \frac{M_0}{M_{1,0}} \sqrt{\frac{1-C_0^2}{1-4\beta_0^2}}
\frac{{\bf L}_0 \!\times\!{\bf b} }{L_ob}
  \label{eq:40}
\end{eqnarray}
If $\beta_0<1/2$ then the
solution is overdamped and Eq.~\ref{eq:37} provides a clear definition.
On the other hand, if $\beta_0>1/2$ then the solution is
underdamped and oscillatory as
\begin{eqnarray}
{\bf C}_\perp(M_1) &=& \left ( \frac{M M_1}{M_0 M_{1,0}} \right )^{1/2}
\times \nonumber \\*
& & ~~ 
\biggr \{ {\bf C}_{\perp,0} \cos \left [ \beta \ln \left (
      \frac{M_1 M_0}{M_{1,0} M} \right ) \right ] + 
\label{eq:41}
\nonumber \\*
& & \left (\frac{\beta_0}{\beta} \frac{M_0}{M_{1,0}} \sqrt{1-C_0^2}
\frac{{\bf L}_0 \!\times\!{\bf b} }{L_ob}
  - \frac{M_0+M_{1,0}}{2\beta  M_2} {\bf C}_{\perp,0} \right ) 
\times
\nonumber \\*
& &
\sin \left [ \beta \ln \left (
      \frac{M_1 M_0}{M_{1,0} M} \right ) \right ]
 \biggr \}
\end{eqnarray}
where $\beta=\beta_0 \sqrt{1-1/\left(2\beta_0\right)^2}$.
\section{Results}
\label{sec:results}

The form of the evolution generally depends on the value of $\beta_0$
specifically whether it exceeds one-half, and the initial and final masses
of the objects in the binary.  It is useful to obtain a more accurate
value of $\beta_0$ for various situations.  Specifically, the rocket
equation, Eq.~\ref{eq:4},
gives
\begin{equation}
\Delta v_{\rm single} = b v_{\rm wind} \ln \frac{M_{1,0}}{M_{1,WD}}
\label{eq:42}
\end{equation}
where $\Delta v_{\rm single}$ is the magnitude of the velocity kick
that the star would have received if it were solitary so
\begin{eqnarray}
\beta_0 &=& \frac{3}{2} \frac{\Delta v_{\rm single}}{\ln
  \frac{M_{1,0}}{M_{1,WD}}}  \sqrt{\frac{Ma}{GM_2^2} }
\label{eq:43}\\
&\approx& 
 0.1  \frac{\Delta v_{\rm single}}{1~{\rm km~
      s}^{-1}} 
\left ( \frac{M}{2 \msun} \frac{a}{1~{\rm AU}} \right )^{1/2}
\left ( \frac{M_2}{1~\msun}  \right )^{-1}
\label{eq:44}\\
&\approx& 
80 \frac{\Delta v_{\rm single}}{1~{\rm km~
      s}^{-1}}
\left ( \frac{M}{1~\msun} \frac{a}{1~{\rm AU}} \right )^{1/2}
\left ( \frac{M_2}{1~\mjup}  \right )^{-1}
\label{eq:45}\end{eqnarray}
where $\mjup$ is the mass of Jupiter and
\begin{equation}
\ln \frac{M_1}{M_{1,WD}} = \ln \frac{M_{1,0}}{0.38 {\rm M}_\odot +
  0.15 M_{1,0}}
\label{eq:46}\end{equation}
was taken to be 0.63, the value appropriate for one solar mass using
the initial-final mass relation of \citet{1983ARA&A..21..271I}.

\subsection{Weak winds}
\label{sec:weak-winds}

Although it might not be obvious from Eq.~(\ref{eq:37}), in the limit
where the wind asymmetry ($\bf b$) vanishes, the eccentricity vector
is constant.  Specifically, 
in the limit of $\beta_0 \ll 1/2$, Eq.~(\ref{eq:37})
through~(\ref{eq:40}) become
\begin{equation}
{\bf C}_\perp(M_1) \approx
{\bf C}_{\perp,0} - 
\beta_0
\left ( 1 - \frac{M_1}{M_{1,0}} \right ) \frac{M_2}{M_{1,0}}
\sqrt{1-C_0^2}
\frac{{\bf L}_0 \!\times\!{\bf b} }{L_ob}
  \label{eq:47}
\end{equation}
so in the limit where the asymmetry of the wind vanishes, the
eccentricity vector remains constant.  In general the eccentricity
will either increase or decrease slightly, but if one considers that
the wind will propel the giant star in a random direction compared to
the original, orbital angular momentum, the final eccentricity is on
average the sum in quadrature of the initial eccentricity and an
increment due to the wind, therefore, slightly larger than the initial
eccentricity.  Furthermore, because on average the two sum in
quadrature, the change in the eccentricity decreases as the initial
eccentricity increases, so it would be difficult for such a weak wind
to unbind a binary.

\subsection{Planetary systems}
\label{sec:planetary-systems}

In the case of a planetary companion, the value of $\beta_0$ is large and
approximately equal to $\beta$ and $M\approx M_1$ which allows
further approximations.  An important quantity in this analysis is the
product of the undamped frequency and the mass of the secondary
\begin{equation}
\beta_0 M_2 = 
0.08 \msun \frac{\Delta v_{\rm single}}{1~{\rm km~
      s}^{-1}}
\left ( \frac{M}{1~\msun} \frac{a}{1~{\rm AU}} \right )^{1/2}
  \label{eq:48}
\end{equation}
which is small compared to the mass lost if the velocity kick is small.
An expansion in $\beta_0 M_2$ for $M_2\ll M_1$, yields
\begin{equation}
{\bf C}_\perp(M_1) \approx
{\bf C}_{\perp,0} - 
\beta_0
\left ( 1 - \frac{M_1}{M_{1,0}} \right ) \frac{M_2}{M_{1,0}}
\sqrt{1-C_0^2}
\frac{{\bf L}_0 \!\times\!{\bf b} }{L_ob}
  \label{eq:49}
\end{equation}
which is identical to the weak wind case, Eq.~(\ref{eq:47}) even
though the undamped frequency is large.  In both cases, the important
factor $\beta_0 M_2$ is small yielding a similar expansion.
Specifically if the initial eccentricity is low, as one would expect
for planetary systems, the final eccentricity of the system will
increase linearly with the strength of the wind and the amount of mass
loss, resulting in final eccentricities of a few percent.   In general
the initial eccentricity and the increment due to the asymmetric wind will add
in quadrature, resulting in a modest gain in eccentricity due to the
wind.

\subsection{Strong winds}
\label{sec:strong-winds}

In the case of a strong wind (a large value of $\Delta v_{\rm single}$
compared to the typical orbital velocities), the value of $\beta_0
M_2$ is large, as is the value of $\beta_0$.  Again $\beta_0$ is
approximately equal to $\beta$, but the two terms that multiply the
sine function in Eq.~(\ref{eq:41}) are similar in magnitude.  The
results of \citet{2007Davis} and \citet{Heyl07kickgc} indicate that
the kicks that white dwarfs receive as giants could be as large as 5-10
km/s, so $\beta_0 M_2 \approx 1 \msun$, so the approximations in
\S\ref{sec:weak-winds} and~\ref{sec:planetary-systems} do not apply.

\begin{figure}[t] 
\includegraphics[width=3.5in]{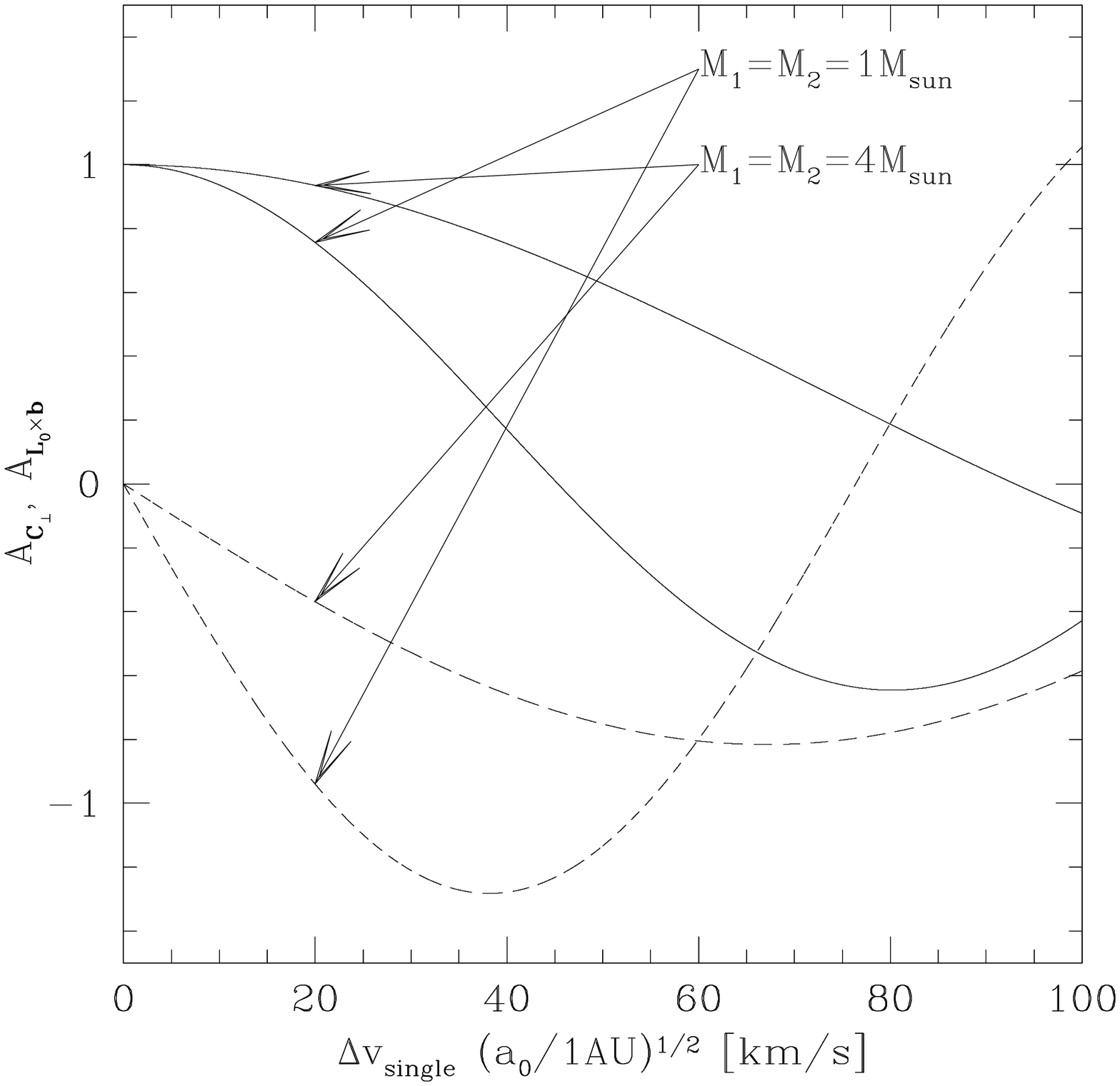} \\
\includegraphics[width=3.5in]{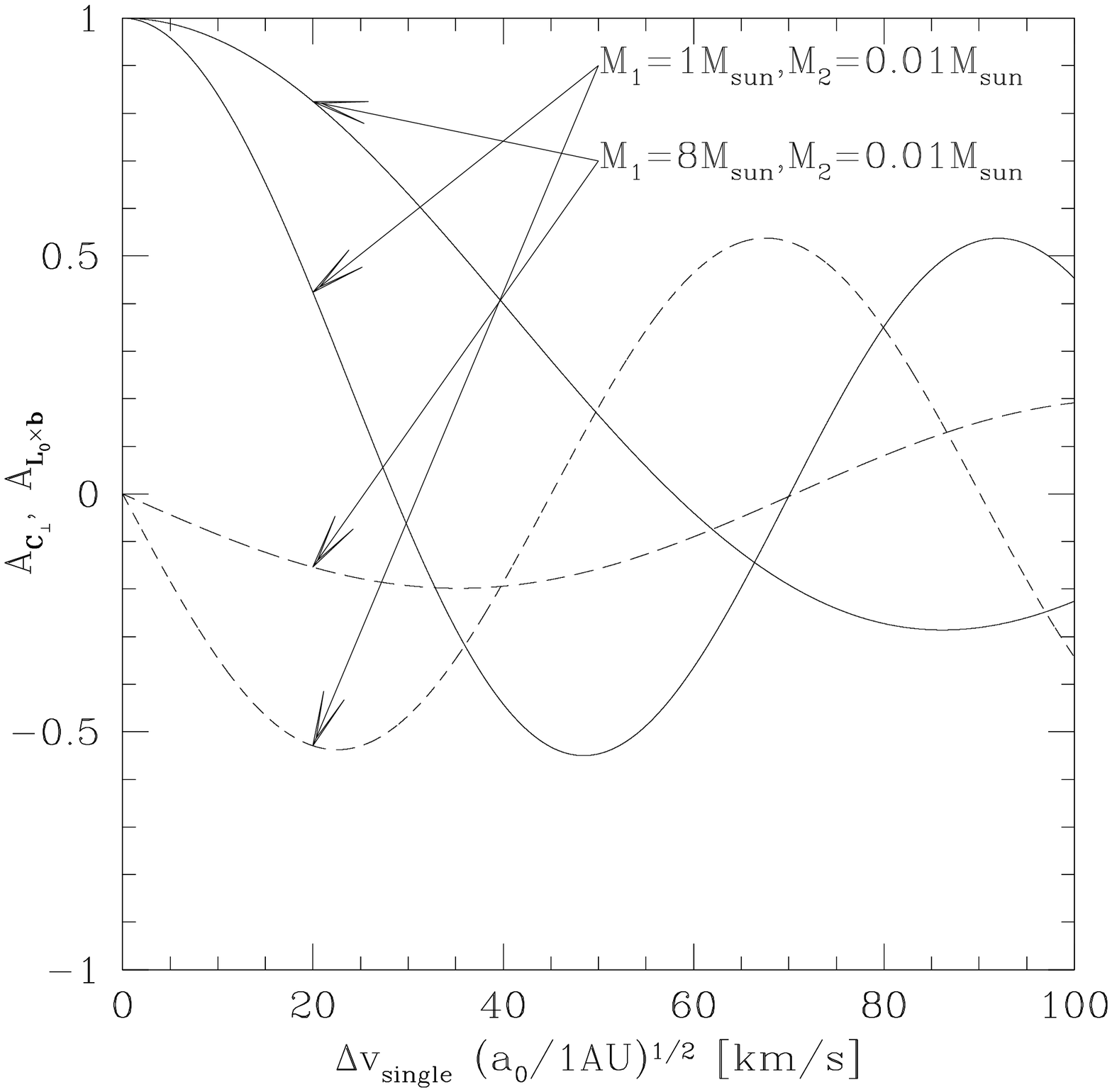}
\caption{The final eccentricity as a function of strength of the wind
  asymmetry and the mass of the primary and secondary.  The upper panel
gives the case for equal-mass binaries, and the lower panel shows the
results for secondaries that are much smaller than the primary (the
results for even smaller secondaries look similar).  The solid curves
follow the component proportional to $C_{\perp,0}$ and the dashed
curves follow the component in the direction of ${\bf L}_0 \times {\bf
b}$. The final value of the eccentricity is given by $e=\left
|{\bf C}_\|+A_{{\bf C}_{\perp,0}} {\bf C}_{\perp,0} + A_{{\bf L}_0\times {\bf
    b}} (L_0 b)^{-1} {\bf L}_0 \times {\bf
b}\right |$}
\label{fig:sine} 
\end{figure}

As before the final Runge-Lenz vector is a linear combination of the
initial Runge-Lenz vector and a vector of length about unity in the
direction of the cross-product of the initial angular momentum and the
asymmetry.  In general the results depicted in Fig.~\ref{fig:sine} are
rather complicated.  
The final Runge-Lenz vector is given by
\begin{equation}
  C_\| = C_{\|,0}, {\bf C}_{\perp} + A_{{\bf C}_\perp} {\bf C}_{\perp,0}
  +  A_{{\bf L}_0 \times {\bf b}} \frac{{\bf L}_0 \times {\bf b}}{L_0 b}
\label{eq:50}
\end{equation}
where the two coefficients are depicted in Fig.~\ref{fig:sine}.

The two contributions are about ninety degrees out of phase, so
the final eccentricity may have little memory of its initial value.
Furthermore, even if the initial eccentricity was small, the second
term is of order unity, so the final eccentricity may be large.  
For a particular white dwarf binary one can infer the initial mass
of the white dwarf from its current mass using, for example,
Eq.~(\ref{eq:46}) and backtrack to obtain the initial semimajor axis
and estimate the contribution to the final eccentricity from kicks of
different magnitudes using Eq.~(\ref{eq:41}).

\subsection{Disruption}

\begin{figure}
\includegraphics[width=3.5in]{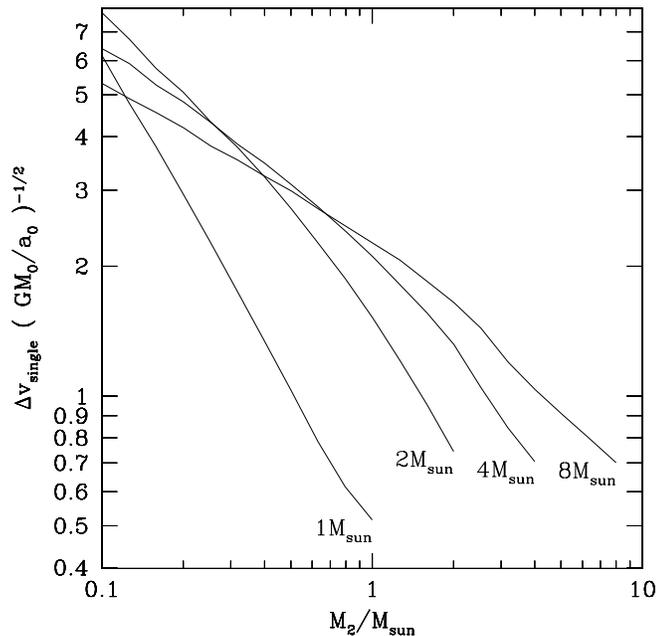}
\caption{The required velocity kick imparted to a single star required
  to disrupt an initially circular binary as a function of the masses
  of the two stars in the system.  The curves are labeled with the
  mass of the primary.  The required velocity kick increases as the
  mass of the secondary decreases because the contribution of the wind
  to the equation of motion of the binary is proportional to $M_2$
  from Eq.~(\ref{eq:1}).  This is a result of the separation of the
  system into the motion of the centre of mass and the motion about
  the centre of mass.  }
\label{fig:distrupt}
\end{figure}

In Fig.~\ref{fig:sine} the length of the final eccentricity vector may
exceed unity; this implies that the binary has become unbound.  To
estimate the velocity kick required to unbind binaries with various
initial properties the maximum eccentricity of an initially circular
orbit is determined.  For small velocity kicks, this typically occurs
at the end of the mass loss, but for larger kicks the peak
eccentricity according to Eq.~(\ref{eq:41}) may be reached while the
mass loss is ongoing.  Fig.~\ref{fig:distrupt} depicts the 
velocity kick that the primary would receive if it were single that
is large enough to unbind the system.
For all of the primary masses, the required velocity kick increases
with decreasing mass of the secondary.  This is simply due to the
translation into the centre of mass frame of the system.  For large
primary masses, the primary loses approximately 85\% of its mass
independent of its mass, so the results become nearly scale-free in
this regime, the required velocity kick is about $\sqrt{GM_0/(2a_0)}$
for equal mass binaries and increases in inverse proportion to the
mass of the secondary.  The dependence becomes shallower when the
final mass of the primary exceeds that of the secondary.

\section{Conclusions}
\label{sec:conclusions}

The possibility of asymmetric winds in asymptotic giant stars opens a
range of new phenomena in the evolution of binary stars.  The presence
of an asymmetry in the wind typically increases the eccentricity of
the binary, so a comparison of the properties of a binaries
containing a white dwarf to binaries of main-sequence stars could
provide an independent probe of the importance of white dwarf kicks.
In equal mass binaries if the wind is sufficiently asymmetric to
induce a kick on the order of the orbital velocity of a single star,
the kick is likely to unbind the system, in contrast with the
symmetric mass loss that only results in an increase in the binary
separation, so the binary fraction of main-sequence stars and white
dwarfs may differ providing an additional probe of this process.  The
effects of white dwarf kicks in planetary systems are more subtle.
Even large kicks do not typically unbind the system; however, kicks
may pump the eccentricity of the orbits to high enough values to
change the resonance structure of multiple planet systems \citep[see
also][]{2002ApJ...572..556D}, possibly resulting in the ejection of
smaller planets from the system.  Finally the white dwarf kicks could
play an important role in the binary evolution and especially double
degenerate binaries, yielding systems that would otherwise be
difficult to form.

\section*{Acknowledgments}

I would like to thank Harvey Richer and Saul Davis for useful
discussions and Camp Byng where some of the work was completed.  The
Natural Sciences and Engineering Research Council of Canada, Canadian
Foundation for Innovation and the British Columbia Knowledge
Development Fund supported this work.  Correspondence and requests for
materials should be addressed to heyl@phas.ubc.ca.  This research has
made use of NASA's Astrophysics Data System Bibliographic Services

\bibliographystyle{mn2e}
\bibliography{mine,wd,physics,math}
\label{lastpage}
\end{document}